\documentclass[12pt]{article}
\usepackage{booktabs}
\usepackage{threeparttable}
\usepackage[colorlinks,bookmarksopen,bookmarksnumbered,citecolor=blue,urlcolor=blue,linkcolor=blue]{hyperref}
\usepackage{float}
\usepackage{amsmath}
\usepackage{graphicx}
\usepackage{subfigure}
\graphicspath{{Pictures/}}
\usepackage{comment}

\usepackage{pifont}

\newcommand{\argmin}{\ensuremath{\operatornamewithlimits{argmin}}}

\def\spacingset#1{\renewcommand{\baselinestretch}%
	{#1}\small\normalsize} \spacingset{1}

\usepackage{amsthm,amsmath,enumerate,amsbsy,amsfonts,amssymb,mathabx,amscd,graphicx,algorithm, indentfirst}
\usepackage{natbib}
\usepackage{geometry}
\usepackage{authblk, caption} %subcaption}
\usepackage{mathrsfs}
\usepackage{epsfig}

\newtheorem{theorem}{Theorem}

\newtheorem{proposition}{Proposition}
\newtheorem{remark}{Remark}

\newtheorem{condition}{Condition}
\newtheorem{assumption}{Assumption}

\usepackage{tablefootnote}
\usepackage{multirow}
\usepackage{xr}

\makeatletter
\newcommand*{\addFileDependency}[1]{% argument=file name and extension
  \typeout{(#1)}
  \@addtofilelist{#1}
  \IfFileExists{#1}{}{\typeout{No file #1.}}
}
\makeatother

\newcommand{\bLambda}{\boldsymbol \Lambda}

\newcommand{\bSigma}{\boldsymbol \Sigma}

\newcommand{\bGamma}{\boldsymbol \Gamma}

\newcommand{\btau}{\boldsymbol \tau}

\newcommand{\bPhi}{\boldsymbol \Phi}

\newcommand{\btheta}{\boldsymbol \theta}
\newcommand{\bgamma}{\boldsymbol \gamma}

\newcommand{\bbeta}{\boldsymbol \beta}

\newcommand{\X}{{\mathbf X}}
\newcommand{\E}{{\mathbf E}}
\newcommand{\A}{{\mathbf A}}
\newcommand{\B}{{\mathbf B}}
\newcommand{\W}{{\mathbf W}}
\newcommand{\I}{{\mathbf I}}

\newcommand{\bA}{{\bf A}}
\newcommand{\bB}{{\bf B}}

\newcommand{\bX}{{\bf X}}

\newcommand{\bQ}{{\bf Q}}

\newcommand{\eR}{\mathbb{R}}

\newcommand{\cov}{\text{Cov}}

\DeclareMathOperator{\vect}{vec}

\makeatletter
\newcommand*{\rom}[1]{\expandafter\@slowromancap\romannumeral #1@}
\makeatother

\makeatletter
\DeclareRobustCommand\widecheck[1]{{\mathpalette\@widecheck{#1}}}
\def\@widecheck#1#2{%
	\setbox\z@\hbox{\m@th$#1#2$}%
	\setbox\tw@\hbox{\m@th$#1%
		\widehat{%
			\vrule\@width\z@\@height\ht\z@
			\vrule\@height\z@\@width\wd\z@}$}%
	\dp\tw@-\ht\z@
	\@tempdima\ht\z@ \advance\@tempdima2\ht\tw@ \divide\@tempdima\thr@@
	\setbox\tw@\hbox{%
		\raise\@tempdima\hbox{\scalebox{1}[-1]{\lower\@tempdima\box
				\tw@}}}%
	{\ooalign{\box\tw@ \cr \box\z@}}}
\makeatother
\RequirePackage[colorlinks,citecolor=blue,urlcolor=blue]{hyperref}

\renewcommand{\baselinestretch}{1.7}

\newcommand{\blind}{1}

\begin{document}
	\if1\blind
	{
		\title{\bf \Large
 Two-way Matrix Autoregressive Model with Thresholds
		}
		\author[1]{Cheng Yu}
        \author[1]{Dong Li}
        \author[2]{Xinyu Zhang\footnote{Correspondence author, xinyu-zhang@uiowa.edu}}
        \author[1,3]{Howell Tong}
	    \affil[1]{\it Department of Statistics and Data Science, Tsinghua University, Beijing, China}
        \affil[2]{\it Department of Statistics and Actuarial Science, University of Iowa, Iowa City, USA}
        \affil[3]{\it Department of Statistics, London School of Economics and Political Science, London, UK}
		\setcounter{Maxaffil}{0}
		
		\renewcommand\Affilfont{\itshape\small}
		%\date{\today}
		\date{\vspace{-5ex}}
		\maketitle
		%\vspace{-2cm}
	} \fi
	\if0\blind
	{
		\bigskip
		\bigskip
		\bigskip
		\begin{center}
			{\Large\bf }
		\end{center}
		\medskip
	} \fi

\setcounter{Maxaffil}{0}
\renewcommand\Affilfont{\itshape\small}
\begin{abstract}
Recently, matrix-valued time series data 
have attracted significant attention in the literature with the recognition of threshold nonlinearity representing a significant advance. However, given the fact that a matrix is a two-array structure, it is unfortunate, perhaps even unusual,  for the threshold literature to 
focus on using the same threshold variable for the rows and the columns. In fact, evidence in economic, financial, environmental and other data shows advantages of allowing the possibilities of two different threshold variables (with possibly different threshold parameters for rows and columns), hence the need for a {\bf Two-}way {\bf M}atrix {\bf A}uto{\bf R}egressive model with {\bf T}hresholds (2-MART). Naturally, two threshold variables pose new and perhaps even fierce challenges, which might be the reason behind the adoption of only one threshold variable in the literature up to now.  
In this paper,  we develop a comprehensive methodology for the 2-MART model, by overcoming various challenges. Compared with existing models in the literature, the new model can achieve greater dimension reduction, much better model fitting, more accurate predictions, and more plausible interpretations.  

\end{abstract}


\section{Introduction} \label{sec:intro}
The growing complexity of time series data has driven the need for collection methods that go beyond traditional vector formats. When time series data are observed across two or more categorical dimensions, they inherently adopt a matrix or tensor structure. Examples of such data include macroeconomic time series where rows represent different industrial sectors and columns represent different regions (\citealp{samadi2025matrix});
international trading data, where rows and columns represent different countries for imports and exports (\citealp{chen2022tradenetwork}); the air pollution data, where rows represent different pollutant concentration readings and columns represent different monitoring stations (\citealp{han2024simultaneous}), 
and global stock data, where rows represent different sector indexes and columns represent different stock markets (\citealp{yu2024matrix}); and others.
Notably, distinct types of structures are represented by the columns and rows of the matrix, which are closely interrelated.
This interrelation makes it crucial to take the matrix alignment into account while modeling the dynamic features of matrix-valued time series. An important contribution to this field is made by \cite{chen2021autoregressive}, who proposed the bilinear matrix autoregressive (MAR) model.\footnote{Recently, there has been an emerging body of literature on matrix-valued time series.
Besides the MAR model, the matrix factor model has been proposed by \cite{wang2019factor} to reduce dimensionality and uncover the underlying correlation structures.
Other works on this topic include \cite{yu2022projected}, \cite{chang2023modelling},  \cite{chen2023statistical}, \cite{gao2023two}, \cite{he2023oneway}, \cite{yuan2023two}, \cite{he2024matrix} and references therein.
There are also works studying the time-varying matrix-valued data, such as \cite{chen2024time}, \cite{he2024online} and \cite{zhang2024adaptive}.} This linear model, which will be introduced in detail later, provides reasonable interpretations of the coefficient matrices and reduces the number of parameters compared with the vectorized formulations.
Nonlinearity is crucial in univariate time series for capturing complex patterns, and it becomes even more significant in matrix-valued time series due to the additional layers of interactions between variables. 
In this paper, to handle nonlinear dynamics, we adopt the threshold principle \citep {tong1990nonlinear} and propose a new approach leading to a bi-nonlinear model, known as the two-way matrix autoregressive model with thresholds (2-MART).
This model aims to capture the intricate nonlinearity within rows and columns, thereby offering a deeper understanding of the underlying nonlinear dynamics.
It improves model specification by accommodating dynamics that vary across different regimes while maintaining easy interpretation.

Specifically, \cite{chen2021autoregressive} proposed the MAR model in a bilinear form:
\begin{equation}\label{MAR}
    \X_t = \A\X_{t-1}\B' + \E_t, \quad t = 1,2,\dots, T,  
\end{equation}
where $\X_t$ is the $m \times n$ matrix observation, $\E_t$ is the $m\times n$ matrix error, $\A$ and $\B$ are $m \times m$ and $n\times n$ autoregressive coefficient matrices for row and column, respectively.   Compared with modelling $\operatorname{vec}(\mathbf{X}_t)$ directly by a vector autoregressive model, the MAR model \eqref{MAR} leverages the matrix structure to reduce the number of parameters and enhance interpretability in practical applications. In this context, \cite{Xiao2022reduced} proposed a reduced-rank MAR model for high-dimensional matrix observations, \cite{li2021multi} and \cite{wang2024high} considered the autoregressive model for tensor-valued time series, while \cite{zhang2024additive} transformed the bilinear component into three different additive models. Expanded details of these models can be found in the review article by \cite{tsay2023matrix}.

The MAR model, as a linear process, has the advantage of being easy to understand and apply, but can be almost powerless in the face of nonlinearity. Hence, the need for a nonlinear model for matrix-valued time series is self-evident.
Recent literature has seen a growing interest in modeling and theoretically analyzing non-linear effects for matrix-valued time series. \cite{liu2022identification} studied threshold matrix-variate factor models. For MAR models, \cite{bucci2024smooth} proposed a smooth transition structure to account for gradual changes in the dynamics, while \cite{wu2023mixture} utilized a mixture structure to analyze potential regime shifts. However, a basic assumption of these studies is that the intrinsic structural changes in the rows are the same as in the columns. 

We argue that this assumption is often unrealistic for practical applications.
Given the inherent two-way complexity of matrix-valued time series, it is essential to account for the two-way nonlinear effects. 
The threshold principle, which holds a prominent place in nonlinear time series analysis, is popular for its capability of modeling many nonlinear phenomena and often offers interpretable results in substantive fields.\footnote{Threshold models have extensive applications in many fields since proposed. Some recent applications include studying, Granger causality in neuroscience \citep{aslan2024nonlinear},  effect of copycat suicides by media \citep{lin2021threshold}, mechanism of economic growth \citep{osinska2020modeling},  bank balance sheet analysis in financial econometrics \citep{zhang2022penalized}, claims reserving in actuarial science \citep{siu2022threshold}, impact of income inequality on the environmental Kuznets curve hypothesis \citep{wang2023does}, among others.
}
Moreover, it can accommodate the two-way nonlinear structure by incorporating two threshold variables, one for each mode.
For example, in the portfolio returns example analyzed in Section \ref{sec:eg}, the matrix-valued time series is constructed by grouping the stocks based on the size and value factors. 
To account for the nonlinear characteristics in the financial market, a threshold model is built with threshold variables aligned with these factors, specifically utilizing the small-minus-big (SMB) and high-minus-low (HML) factors, as inspired by the well-known Fama–French three-factor model.
Note that these two factors respectively account for size and value effect.
This formulation achieves superior model forecasting and also provides valuable insights into understanding the stock market dynamics.
As far as we know, in the literature, most threshold models are concerned with a single threshold variable, and there is almost no work that considers multiple threshold variables in such an interrelated manner.
One of the few exceptions is \cite{zhang2024least}, who studied the estimation and inference of the threshold autoregressive model with more than one threshold variable. % ($K$-TAR).
However, they only considered univariate time series. 
In contrast, matrix-valued time series are interrelated in a more complex way, and our approach models the nonlinearity in a two-way manner, which is both innovative and challenging.

In this paper, we introduce the two-way matrix autoregressive model with thresholds, which relaxes 
the invariant assumption on 
the coefficient matrices in the bilinear model, and allows the threshold variable for the rows to differ from that for the columns. 
It captures different regime-shifting behaviors by following the 
unique 
threshold effects for rows and columns separately. Our approach achieves great dimension reduction, resulting in improved model fitting and prediction performance. Additionally, the enhanced interpretability provides deeper insights into the data mechanisms across various fields.
Our model includes several nonlinear models as special cases, and has the potential to generalize to tensor data. 
We propose a carefully designed estimation procedure that leverages the intrinsic matrix structure and the two-way regime-switching framework, and investigate the limiting distribution of the estimator. 
The two estimated thresholds are proved to be asymptotically independent from each other and from the estimated autoregressive parameters.
In summary, the proposed methods are flexible in capturing different threshold effects for rows and columns of matrix-valued time series, thus broadening the potential applications of threshold models.

The rest of the paper is organized as follows. Section \ref{sec:Model} introduces the two-way matrix autoregressive model with thresholds, and discusses the special cases and generalizations. Section \ref{sec:Estimation} discusses the estimation methods for both the autoregressive coefficients and threshold parameters. The asymptotic properties of the least squares estimator are examined in Section \ref{sec:Asymptotic Properties}. 
Simulation results are provided in Section \ref{sec:Simulation}.
Two application examples, one for portfolio returns and the other for air pollutant analysis, are presented in Section \ref{sec:Real Data} to demonstrate the significance of the model in practice.
Section \ref{sec:Conclusion} concludes.
All proofs, more model generalizations, and more numerical details are relegated to the Supplementary Material.

\section{Model}
\label{sec:Model}
Suppose $\X_t$ is an $m \times n$ matrix for $t\in \mathbb{Z}$.
We propose a model with bi-nonlinear threshold effects.
Since the row and column of $\bX_t$ typically represent different features of the data, it is reasonable to assume that threshold effects may also differ between them.
Hence, we consider a model that incorporates separate threshold variables for rows and columns. Specifically, the 2-MART model is defined as
\begin{equation}\label{Two-TMAR}
	\X_t = 
	\left\{
	\begin{aligned}
		&\A_1 \X_{t-1} \B_1' + \E_{t}, \quad\mbox{if}\quad z_{t-1} \le r,  w_{t-1} \le s,  \\
		&\A_2 \X_{t-1} \B_1' + \E_{t}, \quad\mbox{if}\quad z_{t-1} > r,  w_{t-1} \le s,  \\
		&\A_1 \X_{t-1} \B_2' + \E_{t},\quad\mbox{if}\quad z_{t-1} \le r,  w_{t-1} > s,  \\
		&\A_2 \X_{t-1} \B_2' + \E_{t}, \quad\mbox{if}\quad z_{t-1} > r,  w_{t-1} > s,  \\
	\end{aligned}
	\right.
\end{equation}
where $z_{t-1}$ and $w_{t-1}$ are the threshold variables for row and column respectively, with the corresponding threshold parameters $r$ and $s$. 
Here, $\X_t$ is classified into four distinct regimes.
The coefficient matrices $\A_1, \A_2 \in \mathbb{R}^{m \times m}$ reflect row-wise interactions, between which the switching is decided by the value of $z_{t-1}$.
Similarly, $\B_1, \B_2 \in \mathbb{R}^{n \times n}$ represent
column-wise interactions, with the nonlinear effect caused by $w_{t-1}$.
The errors $\E_t$ are i.i.d. random matrices with zero mean and covariance matrix $\cov(\vect(\E_t))=\bSigma$.

It could also be written in a more parsimonious way as 
\[
\X_t=\sum_{i,j=1,2}\A_i \X_{t-1} \B_j'I_{t,i,j}(\btau) + \E_{t},
\]
where $I_{t,i,j}(\btau)$ is the indicator function for $i,j \in \{1,2\}$.
Specifically,
$I_{t,1,1}(\btau)=I(z_{t-1} \le r,  w_{t-1} \le s)$, $I_{t,2,1}(\btau)=I(z_{t-1} > r,  w_{t-1} \le s)$,
$I_{t,1,2}(\btau)=I(z_{t-1} \le r,  w_{t-1} > s)$, and
$I_{t,2,2}(\btau)=I(z_{t-1} > r,  w_{t-1} > s)$.

\begin{remark}\label{rem:hete}
    In (\ref{Two-TMAR}), we assume a model with homogeneous mean and covariance matrix across regimes, primarily to simplify the notation. However, the methods and theory presented in this paper can be easily extended to cases with regime-dependent means or covariance matrices. For further discussions and details on the heterogeneous cases, interested readers can refer to  Section S.6 of the Supplementary Material.
\end{remark}

Hence, the threshold effects are separate in the sense that the row-switching and  column-switching dynamics are affected by different threshold variables, while they  
are also joint and interrelated since the MAR model is a two-way model.
This feature distinguishes our 2-MART model from \cite{zhang2024least}.

By  vectorization,  model  \eqref{Two-TMAR} becomes a threshold vector autoregressive model \citep{tsay1998testing} with 4 regimes and coefficients matrices given by Kronecker products,
\begin{equation}\label{eq:vector}
   \vect(\X_t)=\sum_{i,j=1,2}(\B_j \otimes \A_i) \vect(\X_{t-1})I_{t,i,j}(\btau) + \vect(\E_{t}).
\end{equation}
The coefficients $\B_j \otimes \A_i$ represent row-column interactions in each regime.
Since there are $O(m^2n^2)$ parameters in this vector model for $\vect(\X_t)$, the two-way matrix design in (\ref{Two-TMAR}) and (\ref{eq:vector}) reduces the number of parameters to $O(m^2+n^2)$,
which represents a substantial reduction for large $mn$.

As for the identification, note that the Kronecker products $\B_j \otimes \A_i$ for $1\le i, j \le 2$ remain unchanged 
if $\B_1$ and $\B_2$ are multiplied by a non-zero constant $c$ while $\A_1$ and $\A_2$ are divided by the same constant. Thus, we impose the identification condition that $\|\A_1 \| = 1$, where $\|\cdot\|$ denotes the Frobenius norm (Euclidean norm),
and $(\B_1)_{1,1} \geq 0$, i.e., the first diagonal element of $\B_1$ should be non-negative.

Our model allows for different threshold variables for rows and columns, encompassing simpler models as special cases.
A notable special case occurs when $z_t = w_t$ but $r \neq s$, resulting in a single threshold variable with different threshold parameters for rows and columns. 
This is related to the multiple-regime threshold autoregressive model \citep{li2012least} but with a two-way matrix structure. 
Specifically, we have $3$ regimes as follows:
\begin{equation}\label{Multiple-TMAR}
	\X_t = 
	\left\{
	\begin{aligned}
		&\A_1 \X_{t-1} \B_1' + \E_{t}, && \quad\mbox{if} \quad z_{t-1} \le r \land s,  \\
		& \A_2 \X_{t-1} \B_1' I(r<s) + \A_1 \X_{t-1} \B_2' I(s<r)+ \E_{t}, &&\quad\mbox{if} \quad  r \land s < z_{t-1} \le r \lor s,  \\
		&\A_2 \X_{t-1} \B_2' + \E_{t}, &&\quad\mbox{if} \quad z_{t-1} > r \lor s,  \\
	\end{aligned}
	\right.
\end{equation}
where $r \land s=\min\{r, s\}$
and $r \lor s=\max\{r, s\}$.
It indicates that regime switching for row and column is triggered by the same threshold variable $z_{t-1}$, but at different levels (row at $r$ and column at $s$).
The second regime serves as an intermediate zone between the first and third regimes, softening the hard threshold border between the two regimes and enhancing the model's flexibility.
We note that this structure is also novel in the matrix-valued time series literature, and we denote it as the Softened Matrix Autoregressive with Thresholds (SMART) model.
It proves particularly effective in modeling temperature's nonlinear effects on pollutants across sites;  see Section \ref{sec:Application2-Pollution}.

The threshold MAR model is another special case of our model with $z_t = w_t$ and $r = s$, i.e., with a single threshold variable and a single threshold parameter. Specifically,
\begin{equation}\label{TMAR}
	\X_t = 
	\left\{
	\begin{aligned}
		&\A_1 \X_{t-1} \B_1' + \E_{t}, \quad\mbox{if} \quad z_{t-1} \le r, \\
		&\A_2 \X_{t-1} \B_2' + \E_{t}, \quad\mbox{if} \quad z_{t-1} > r. \\
	\end{aligned}
	\right.
\end{equation}
This model indicates that the regime switching for rows and columns is triggered by the same threshold variable $z_{t-1}$ and simultaneously at level $r$.
This is a natural and straightforward application of the threshold principle in the context of the MAR model.
However, it does not incorporate separate nonlinear effects for rows and columns, making it less flexible compared with models \eqref{Two-TMAR} and \eqref{Multiple-TMAR}, as evidenced by the empirical results in Section \ref{sec:Real Data}.

\begin{remark} \label{rem:tensor}
Our model for the matrix-valued time series can be generalized to higher-order data.
For the tensor-valued time series $\mathcal{X}_t \in \mathcal{R}^{N_1 \times \cdots \times N_K}$, a $K$-way tensor autoregressive model with thresholds can be built. Specifically, 
\begin{align}
\begin{split}\label{KWTTAR}
    \mathcal{X}_t = \mathcal{X}_{t-1} &\times_1 \big(\mathbf{A}_1^{(1)} I(z_{1,t-1} \le r_1) + \mathbf{A}_2^{(1)}  I(z_{1,t-1} > r_1) \big) \times_2 \cdots \\
    &\times_{K} \big(\mathbf{A}_1^{(K)} I(z_{K,t-1} \le r_K) 
    + \mathbf{A}_2^{(K)}  I(z_{K,t-1} > r_K) \big) 
    + \mathcal{E}_t,
\end{split}
\end{align}
where for $k = 1, \dots, K$, $z_{k,t-1}$ is the threshold variable for mode-$k$ of the tensor time series $\mathcal{X}_t$ with threshold parameter $r_k$, $\mathbf{A}_{1}^{(k)}$ and $\mathbf{A}_{2}^{(k)}$ are $N_k \times N_k$ mode-$k$ autoregressive matrices in  regimes decided by $z_{k,t-1}$, and $\mathcal{E}_t \in \mathcal{R}^{N_1\times \dots \times N_K}$ is the tensor-valued noise. Clearly, when the number of mode $K=2$, the model \eqref{KWTTAR} reduces to our 2-MART model \eqref{Two-TMAR}. 
The estimation and inference of tensor models are more complex due to the high-dimensional nature and intricate interactions within the data. These challenges necessitate careful exploration and potentially specialized algorithms. Therefore, we leave these aspects for future work.
\end{remark}

In summary, we propose a general framework \eqref{Two-TMAR} for studying the two-way nonlinear effects in matrix-valued time series, including special cases such as the SMART model \eqref{Multiple-TMAR} and threshold MAR model \eqref{TMAR}, as well as potential generalizations to the heterogeneous case (Remark \ref{rem:hete}) and the tensor case (Remark \ref{rem:tensor}). Given the scope of this paper, we focus on the model \eqref{Two-TMAR} to provide a thorough analysis without overextending our study, examining its estimation, inference, and empirical properties.

\section{Estimation}
\label{sec:Estimation}
Denote $\bbeta=(\vect(\A_1)', \vect(\A_2)', \vect(\B_1)', \vect(\B_2)')'$, $\btau=(r,s)'$, and
$\btheta = (\bbeta', \btau')' \in \eR^{2(m^2 + n^2) + 2}$. 
Let the true parameter be $\btheta_0 =(\bbeta_0', \btau_0')' = (\vect(\A_{10})', \vect(\A_{20})', \vect(\B_{10})', \vect(\B_{20})',$ $r_0, s_0)'$. Suppose that the observations
$\{\X_1,...,\X_T\}$ of length $T$ are from the 2-MART model \eqref{Two-TMAR} with the true parameter $\btheta_0$.
We propose a least squares estimator (LSE) for model\eqref{Two-TMAR}. Specifically, 
the sum-of-squared-error function $L_T(\btheta)$ is defined as
\begin{align}\label{LS_loss}
\begin{split}
L_T(\btheta) 
&= \frac{1}{T}\sum_{t=2}^{T}\Big\|\X_t - \sum_{i,j=1}^2 \A_i\X_{t-1}\B_j'I_{t,i,j}(\btau) \Big\|^2\\
&= \frac{1}{T}\sum_{t=2}^{T}\Big\|\vect(\X_t) - \sum_{i,j=1}^2 (\B_j \otimes \A_i)\vect(\X_{t-1})I_{t,i,j}(\btau)\Big\|^2.
\end{split}
\end{align}
The minimizer of $L_T(\btheta)$ is called the LSE of $\btheta_0$, i.e.,
\begin{equation*}%\label{LSE}
    \widehat{\btheta} = \argmin_{\btheta \in \Theta} L_T(\btheta).
\end{equation*}

Given a threshold parameter $\btau$, by taking partial derivatives of \eqref{LS_loss} with respect to entries of $\mathbf{A}_i$ and $\mathbf{B}_{j}$ for $i, j \in \{1,2\}$ respectively, we obtain the gradient condition for the LSE as
\begin{align}\label{eq:grad}
\begin{split}
    \sum_{j}\sum_{t}\mathbf{A}_{i}\mathbf{X}_{t-1}\mathbf{B}_{j}'\mathbf{B}_{j}\mathbf{X}_{t-1}' I_{t,i,j}(\btau) - \sum_{j}\sum_{t}\mathbf{X}_{t}\mathbf{B}_{j}\mathbf{X}_{t-1}'I_{t,i,j}(\btau) &= \mathbf{0}, \, i=1,2.\\
    \sum_{i}\sum_{t}\mathbf{B}_{j}\mathbf{X}_{t-1}'\mathbf{A}_{i}'\mathbf{A}_{i}\mathbf{X}_{t-1} I_{t,i,j}(\btau) - \sum_{i}\sum_{t}\mathbf{X}_{t}'\mathbf{A}_{i}\mathbf{X}_{t-1}I_{t,i,j}(\btau) &= \mathbf{0}, \, j=1,2,
\end{split}
\end{align}
where $\sum_{i}$ and $\sum_{j}$ respectively represent $\sum_{i=1}^2$ and $\sum_{j=1}^2$, and $\sum_{t}$ represents $\sum_{t=2}^T$.
Then, we iteratively update the estimates for  
$\mathbf{A}_1$ and $\A_2$ ($\mathbf{B}_1$ and $\B_2$) while keeping $\mathbf{B}_1$ and $\B_2$ ($\mathbf{A}_1$ and $\A_2$) fixed. Specifically, we update $\mathbf{B}_1$ and $\B_2$ given $\mathbf{A}_1$ and $\A_2$ as 
\begin{equation*}
    \mathbf{B}_{j} \longleftarrow \Big(\sum_{i}\sum_{t} \mathbf{X}_{t}'\mathbf{A}_{i}\mathbf{X}_{t-1} I_{t,i,j}(\btau) \Big) \Big( \sum_{i}\sum_{t}\mathbf{X}_{t-1}'\mathbf{A}_{i}'\mathbf{A}_{i}\mathbf{X}_{t-1} I_{t,i,j}(\btau) \Big)^{-1}, \, j = 1,2,
\end{equation*}
and similarly the iteration of updating $\A_1$ and $\A_2$ given $\B_1$ and $\B_2$ is
\begin{equation*}
    \mathbf{A}_{i} \longleftarrow \Big(\sum_{j}\sum_{t} \mathbf{X}_{t}\mathbf{B}_{j}\mathbf{X}_{t-1}' I_{t,i,j}(\btau) \Big) \Big( \sum_{j}\sum_{t}\mathbf{X}_{t-1}\mathbf{B}_{j}'\mathbf{B}_{j}\mathbf{X}_{t-1}' I_{t,i,j}(\btau) \Big)^{-1}, \, i = 1,2.
\end{equation*}

We denote the iterative  least squares estimator as $\widehat{\mathbf{A}}_i$ and $\widehat{\mathbf{B}}_j$. The iterative least squares method may converge to a local minimum. In practice, we recommend using the least squares estimator of the MAR model in \cite{chen2021autoregressive} as the initial values for the iteration, 
which is the solution of the least squares problem
$\min_{\bA, \bB}\sum_{t}\|\mathbf{X}_t - \bA\mathbf{X}_{t-1}\bB' \|^{2}$ that ignores the threshold effect. This option demonstrates good performance in our simulations. 
We follow the identification condition and renormalize in each iteration to ensure that $\|\widehat{\mathbf{A}}_1 \| = 1$ and $(\widehat{\mathbf{B}}_{1})_{1,1}\ge 0$.

For the threshold parameters $\btau$, given that $L_{T}(\btheta)$ is discontinuous in $\btau$, we employ the multi-parameter grid-search algorithm to obtain the estimator $\widehat{\btau}$.
Specifically, for each fix parameter $\btau$, we can calculate the LSE  $\widehat{\bbeta}(\btau)$ and the corresponding minimal value  $\widetilde{L}_{T}(\btau) = L_{T}(\widehat{\bbeta}(\btau), \btau)$. Since $\widetilde{L}_{T}(\btau)$ takes only finitely many values, the minimizer $\widehat{\btau}$ of $\widetilde{L}(\btau)$ can be determined through grid search.
Hence, the LSE is obtained as $\widehat{\btheta} = (\widehat{\bbeta}', \widehat{\btau}')'$ with $\widehat{\bbeta} = \widehat{\bbeta}(\widehat{\btau})$.

\section{Asymptotic Properties}
\label{sec:Asymptotic Properties}

In this section, we derive probabilistic properties of the 2-MART model \eqref{Two-TMAR}, and the asymptotics for its LSE.
We first introduce several assumptions.
\begin{assumption}\label{ass:et}
$\{\E_t\}$ is i.i.d. with zero mean and finite variance. The covariance matrix $\bSigma$ of $\vect(\E_{t})$ is non-singular.
Its density $f_{\epsilon}(\cdot)$ is bounded, continuous, and positive on $\mathbb{R}^{m \times n}$.
\end{assumption}
\begin{assumption}\label{ass:theta}
The parameter space $\Theta$ is a compact subset of $\eR^{2(m^2 + n^2) + 2}$.
The coefficients $\A_{10},\A_{20},\B_{10}$ and $\B_{20}$ are non-singular.
\end{assumption}
\begin{assumption}\label{ass:zw}
Let $\{(\vect(\X_t)',z_t,w_t)'\}$ be strictly stationary and ergodic,
where $\{(z_t,w_t)'\}$ are random vectors with a bounded, continuous, and positive density $\pi(\cdot,\cdot)$ on $\mathbb{R}^2$.
Denote the marginal density of $z_t$ and $w_t$ as $\pi_1(\cdot)$ and $\pi_2(\cdot)$, respectively.
If $\{(z_t,w_t)'\}$ are exogenous, we further assume they are Markovian.
\end{assumption}

Assumption \ref{ass:et} is standard in the threshold literature.
The stationarity and ergodicity in Assumption \ref{ass:zw} are intricate, and we provide a sufficient condition for the special case when $z_{t-1}$ and $w_{t-1}$ are endogenous.
The choices of endogenous variables include functions of lagged $\X_{t}$.
\begin{proposition}\label{prop}
Let $\|\cdot\|_2$ denote the spectral norm.
    When $z_t$ and $w_t$ are both endogenous, if $\max_{i,j=1,2}\|\A_i\|_2 \|\B_j\|_2<1$ and Assumption \ref{ass:et} hold, then $\X_t$ is stationary and ergodic.
\end{proposition}
Now, we have the following theorem on the consistency of $\widehat{\btheta}$. 

\begin{theorem}\label{thm:consistency}
Suppose Assumptions \ref{ass:et}-\ref{ass:zw} hold. 
We further assume (i) $\mathbb{E}\{(\X_t)_{i,j}^2\}<\infty$ for $1\leq i \leq m$ and $1\leq j \leq n$;
(ii) $\A_{10} \neq \A_{20}$ and $\B_{10} \neq \B_{20}$.
Then, $\widehat{\btheta} \to {\btheta}$ a.s. as $T \to \infty$.
\end{theorem}

We further introduce several assumptions as follows.
\begin{assumption}\label{ass:fg}
Suppose $\mathbb{E}\{(\E_t)_{i,j}^4+(\X_t)_{i,j}^4\}<\infty$ for $1\leq i \leq m$ and $1\leq j \leq n$.
Let $f(r)=\mathbb{E}(\mathcal{X} | z_t=r)$ and $g(s)=\mathbb{E}(\mathcal{X} | w_t=s)$ with $\mathcal{X}$ being a fixed random variable in the set
$ \left\{\|\X_t\|^3, |w_t|+|z_t|, \|\X_t\|^2 (|z_{t}|+|w_{t}|),\|\E_{t+1}\|_{\max} \|\X_t\|_{\max} (|z_{t+1}|+|w_{t+1}|),\|\X_t \X_t'\|\right\}$, where $\|\cdot\|_{\max}$ denotes the maximum absolute value of the entries in matrix.
For any $\mathcal{X}$ in the set, suppose $f(r)$ is continuous at $r_0$, and $g(s)$ at $s_0$.
\end{assumption}

Let $\mathbf{y}_t=(\vect(\X_t)', z_t, w_t)'$. 
Then $\{\mathbf{y}_t: t\geq 0\}$ is automatically a Markov chain with respect to its natural filtration.
Denote its $k$-step transition probability by $\mathbf{P}^k(\mathbf{y},A)$,
where $\mathbf{y}\in \mathbb{R}^{mn+2}$ and $A$ is a Borel set of $\mathbb{R}^{mn+2}$.

\begin{assumption}\label{ass:markov}
$\{\mathbf{y}_t:t=0,1,...\}$ admits a unique invariant measure $\Pi(\cdot)$
such that there exist $K>0$ and $0<\rho<1$, 
for any $\mathbf{y} \in \mathbb{R}^{mn+2}$ and any integer $k \geq 1$, $\|\mathbf{P}^k(\mathbf{y},\cdot)-\Pi(\cdot)\|_\mathrm{v}\leq K\rho^k(1+\|\mathbf{y}\|)$, 
where $\|\cdot\|_\mathrm{v}$ denotes the total variation norm.
\end{assumption}

\begin{assumption}\label{ass:identify} 
Let $\bGamma=\mathbb{E}(\X_{t}|z_{t}=r_0,w_{t}=s_0)$. Assume that $\|\A_{i0} \bGamma \B_{j0}'-\A_{k0} \bGamma \B_{l0}'\|\neq 0$
for any $i,j,k,l=1,2$ such that $i \neq k$ or  $j \neq l$.
\end{assumption}

Assumption \ref{ass:fg} gives moment and conditional moment conditions, where the latter are mainly for cases with exogenous threshold variables.
Under Assumption \ref{ass:markov}, $\{\mathbf{y}_t\}$ is $V$-uniformly ergodic with $V(\cdot)=K(1+\|\cdot\|)$, a condition which is stronger than geometric ergodicity.
For the concept of $V$-uniform ergodicity, refer to Chapter 16 in \citet{meyn2009markov}.
Assumption \ref{ass:identify} indicates that the autoregressive function is discontinuous at the threshold $(r_0,s_0)$, and the threshold effect is fixed.
The identifiability condition (ii) in Theorem \ref{thm:consistency} is a necessary condition of Assumption \ref{ass:identify}. 

\begin{theorem}\label{thm:rate}
If Assumptions \ref{ass:et}-\ref{ass:identify} hold and $\btheta$ is an interior point of $\Theta$, then
\begin{itemize}
\item[$\mathrm{(i)}$] $T\|\widehat{\btau}-\btau_0\|=O_p(1)$;
\item[$\mathrm{(ii)}$] 
Let $\I_m$ and $\I_n$ be the identity matrix of dimension $m$ and $n$, respectively. 
Define
\begin{align*}
    \W_t'=\Big[ &\sum_{j=1,2}(\B_{j0}\X_{t-1}'I_{t,1,j}(\btau_0)) \otimes \I_m \,\vdots\, \sum_{j=1,2}(\B_{j0}\X_{t-1}'I_{t,2,j}(\btau_0)) \otimes \I_m \\ 
    \,\vdots\, & \I_n \otimes \sum_{i=1,2}(\A_{i0}\X_{t-1}I_{t,i,1}(\btau_0)) 
\,\vdots\, \I_n \otimes \sum_{i=1,2}(\A_{i0}\X_{t-1}I_{t,i,2}(\btau_0))
\Big],
\end{align*}
and $\mathbf{H}=\mathbb{E}(\W_t \W_t')+ \bgamma \bgamma'$ with $\bgamma =(\vect(\A_{10})',\mathbf{0},\mathbf{0},\mathbf{0})'$.
Let $\Xi=\mathbf{H}^{-1}\mathbb{E}(\W_t \bSigma \W_t') \mathbf{H}^{-1}$.
Then,  as $T \to \infty$, it follows that 
\begin{equation*}
\sqrt{T}(\vect(\widehat{\bbeta})-\vect(\bbeta_0))\to_d \mathcal{N}(0,\Xi).
\end{equation*}
\end{itemize}
\end{theorem}

Theorem \ref{thm:rate} establishes the asymptotic Gaussianity of the LSE of the autoregressive coefficients.
Also, it demonstrates that the convergence rate of $\widehat{\btau}$ is $T$, indicating super-efficiency. 
Now, we examine the limiting distribution of $T(\widehat{\btau}-{\btau}_0)$.
Define the following:
\begin{align}\label{eq:gamma}
	\begin{split}
		\gamma_t^{(1)}=&\xi_t\left(\B_{20} \otimes (\A_{20}-\A_{10})\right)I(w_{t-1}>s_0)+\xi_t\left(\B_{10} \otimes (\A_{20}-\A_{10})\right)I(w_{t-1}\leq s_0),\\
		\gamma_t^{(2)}=&\xi_t\left(\B_{20} \otimes (\A_{10}-\A_{20})\right)I(w_{t-1}>s_0)+\xi_t\left(\B_{10} \otimes (\A_{10}-\A_{20})\right)I(w_{t-1}\leq s_0),\\
		\gamma_t^{(3)}=&\xi_t\left((\B_{20}-\B_{10}) \otimes \A_{10}\right)I(z_{t-1}>r_0)+\xi_t\left((\B_{20}-\B_{10}) \otimes \A_{20}\right)I(z_{t-1}\leq r_0),\\
		\gamma_t^{(4)}=&\xi_t\left((\B_{10}-\B_{20}) \otimes \A_{10}\right)I(z_{t-1}>r_0)+\xi_t\left((\B_{10}-\B_{20}) \otimes \A_{20}\right)I(z_{t-1}\leq r_0),
	\end{split}
\end{align}
where $\xi_t(\bPhi)$ is a function defined for any $\bPhi \in \eR^{mn \times mn}$,
\[
\xi_t(\bPhi)=\|\bPhi\vect(\X_{t-1})\|^2
+2\vect(\E_t)'\bPhi\vect(\X_{t-1}).
\]
We introduce two independent one-dimensional two-sided compound Poisson processes,
$\{\mathcal{P}_1(u), u \in \mathbb{R}\}$ and $\{\mathcal{P}_2(v), v \in \mathbb{R}\}$, defined as
\begin{align*}
	\mathcal{P}_1(u)=&I(u>0)\sum_{k=1}^{N_1(u)}\zeta_k^{(1)}
	+I(u\leq 0)\sum_{k=1}^{N_2(-u)}\zeta_k^{(2)},\\
	\mathcal{P}_2(v)=&I(v>0)\sum_{k=1}^{N_3(v)}\zeta_k^{(3)}
	+I(v\leq 0)\sum_{k=1}^{N_4(-v)}\zeta_k^{(4)},
\end{align*}
where $\{N_1(u),u\geq 0\}$ and $\{N_2(u),u\geq 0\}$ are independent Poisson processes with $N_1(0)=N_2(0)=0$ a.s. and the same jump rate $\pi_1(r_0)$. 
The sequence $\{\zeta_k^{(1)}:k\geq 1\}$ are i.i.d. from $F_1(\cdot|r_0)$ and $\{\zeta_k^{(2)}:k\geq 1\}$ from $F_2(\cdot|r_0)$,
and they are mutually independent,
where $F_1(\cdot|r_0)$ is the conditional distribution of $\gamma_2^{(1)}$ given $z_{1}=r_0$,
and $F_2(\cdot|r_0)$ is that of $\gamma_2^{(2)}$. 
In a similar manner, $\{N_3(v),v\geq 0\}$ and $\{N_4(v),v\geq 0\}$ are  independent Poisson processes with $N_3(0)=N_4(0)=0$ a.s. and the same jump rate $\pi_2(s_0)$. 
The sequences $\{\zeta_k^{(3)}:k\geq 1\}$ are i.i.d. from $G_1(\cdot|s_0)$ and $\{\zeta_k^{(4)}:k\geq 1\}$ from $G_2(\cdot|s_0)$,
and they are mutually independent,
where $G_1(\cdot|s_0)$ is the conditional distribution of $\gamma_2^{(3)}$ given $w_{1}=s_0$,
and $G_2(\cdot|s_0)$ that of $\gamma_2^{(4)}$. 
We use the right continuous version for $N_1(u)$ and $N_3(v)$,
and the left continuous version for $N_2(u)$ and $N_4(v)$.

We further define a spatial compound Poisson process
\[\psi(u,v)=\mathcal{P}_1(u)+\mathcal{P}_2(v).\]
Since all jump distributions have positive means by Assumption \ref{ass:identify}, $\psi(u,v)$ clearly goes to $+\infty$ a.s. when $|u|$, $|v|\rightarrow\infty$.
Consequently, there exists a unique 2-dimensional cube $[\boldsymbol{M_-,M_+})\equiv[M_-^{(1)},M_+^{(1)})\times
[M_-^{(2)},M_+^{(2)})$, where $\psi(u,v)$ attains its global minimum a.s.: $[\boldsymbol{M_-,M_+})=\arg\min_{(u,v)\in \mathbb{R}^2}\psi(u,v)$.
Given the independence of $\mathcal{P}_1(u)$ and $\mathcal{P}_2(v)$,
the minimization simplifies to $[M_-^{(1)},M_+^{(1)})=\arg\min_{u \in \mathbb{R}}\mathcal{P}_1(u)$, $[M_-^{(2)},M_+^{(2)})=\arg\min_{v \in \mathbb{R}}\mathcal{P}_2(v)$.
Thus, $M_-^{(1)}$ and $M_-^{(2)}$ are independent. Now, we can state our results.

\begin{theorem}\label{thm:asydist}
If Assumptions \ref{ass:et}--\ref{ass:identify} hold, then $T(\widehat{\btau}-\btau_0)$ converges weakly to $\boldsymbol{M_-}$ and its components are asymptotically independent as $T \to \infty$. 
Further, $T(\widehat{\btau}-\btau_0)$ is asymptotically independent
of $\sqrt{T}({\widehat{\bbeta}}-{\bbeta_0})$.
\end{theorem}

The distribution of $\boldsymbol{M_-}$ lacks a closed form, necessitating the use of numerical methods. This task becomes particularly challenging when the threshold variables are exogenous. To address this, we employ the weighted Nadaraya-Watson approach proposed by \cite{zhang2024least}, as detailed in their Algorithms A and B. This method enables us to calculate robust confidence intervals for the threshold parameters irrespective of whether the threshold variables are endogenous or exogenous.

\section{Simulation Studies}
\label{sec:Simulation}

In this section, we demonstrate the finite sample performance of the LSE, as proposed in Section \ref{sec:Estimation}, across various settings by altering the matrix dimensions $(m,n)$, the length $T$ of observations, and the error covariance matrix $\bSigma$.

The observed data $\mathbf{X}_t$ are generated from the 2-MART model in \eqref{Two-TMAR}
with the $m\times m$ coefficient matrices $\bA_1$ and $\bA_2$ and $n\times n$ coefficient matrices $\bB_1$ and $\bB_2$ as follows,
\begin{align*}
\mathbf{A}_1 = c_1 \begin{bmatrix}
1 & 1 & \cdots & 1 \\
1 & 1 & \cdots & 1 \\
\vdots & \vdots & \ddots & \vdots \\
1 & 1 & \cdots & 1
\end{bmatrix},
\quad \quad
\mathbf{A}_2 = c_2 \begin{bmatrix}
1 & -0.5 & \cdots & -0.5 \\
-0.5 & 1 & \cdots & -0.5 \\
\vdots & \vdots & \ddots & \vdots \\
-0.5 & -0.5 & \cdots & 1
\end{bmatrix},\\
\mathbf{B}_1 = c_3 \begin{bmatrix}
1 & 1 & \cdots & 1 \\
1 & 1 & \cdots & 1 \\
\vdots & \vdots & \ddots & \vdots \\
1 & 1 & \cdots & 1
\end{bmatrix},
\quad \quad
\mathbf{B}_2 = c_4 \begin{bmatrix}
1 & -0.3 & \cdots & -0.3 \\
-0.3 & 1 & \cdots & -0.3 \\
\vdots & \vdots & \ddots & \vdots \\
-0.3 & -0.3 & \cdots & 1
\end{bmatrix},
\end{align*}
where positive constants $c_i$'s are chosen such that $\|\A_1 \| = \|\A_2 \| = 1$  and  $\|\B_1\| = \|\B_2\| = 0.8$ for identifiability and stationarity of the model. 

We consider two settings for the covariance matrix $\bSigma=\operatorname{Cov}(\operatorname{vec}(\mathbf{E}_t))$:
\begin{itemize}
    \item Setting I: \, $\bSigma = \mathbf{I}_{mn}$.
    \item Setting II:  $\bSigma = \bSigma_c \otimes \bSigma_r$,
where $\bSigma_r$ and $\bSigma_c$ represent row-wise and column-wise covariances, respectively.
\end{itemize}
Specifically,
the $m \times m$ matrix $\bSigma_r$ is generated as $\bSigma_r = \bQ \bLambda \bQ^\top$, with $\bQ$ being a random orthonormal matrix and $\bLambda$ being a diagonal matrix whose entries are the absolute values of i.i.d. $\mathcal{N}(0, 1)$ random variables. The $n \times n$ matrix $\bSigma_c$ is generated in a similar manner. 
Motivated by the example in Section \ref{sec:Application1-portfolio}, the threshold variables $z_t$ and $w_t$ are defined as
\begin{equation*}
    z_{t} = \frac{1}{n}\sum_{j=1}^n\{(\mathbf{X}_t)_{m, j} - (\mathbf{X}_t)_{1, j}\}, \quad
    w_{t} = \frac{1}{m}\sum_{i=1}^m\{(\mathbf{X}_t)_{i, n} - (\mathbf{X}_t)_{i, 1}\},
\end{equation*}
with the true threshold parameters $r = 0.02$ and $s = -0.02$. 
Below we will assess the finite-sample performance of the LSE of the autoregressive matrices as well as the threshold parameters.

\begin{figure}[!htbp]
	\centering	\includegraphics[height=9cm, width=15cm]{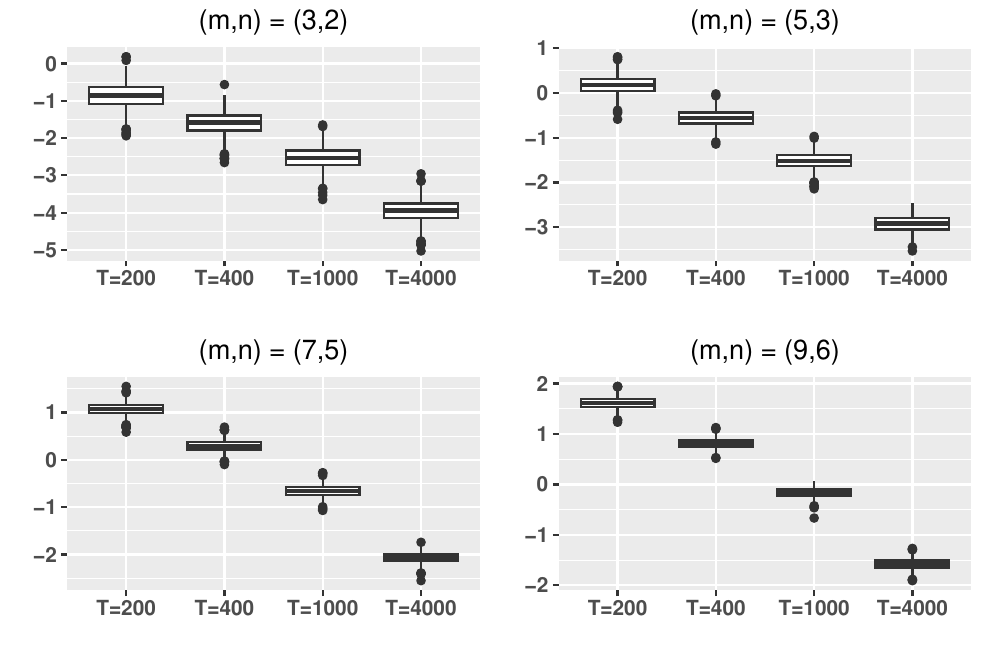}
	\caption{The box plot of the estimation error under Setting I with different $T$ and $(m,n)$.}
	\label{Plot: LS1}
\end{figure}

\begin{figure}[!htbp]
	\centering  \includegraphics[height=9cm, width=15cm]{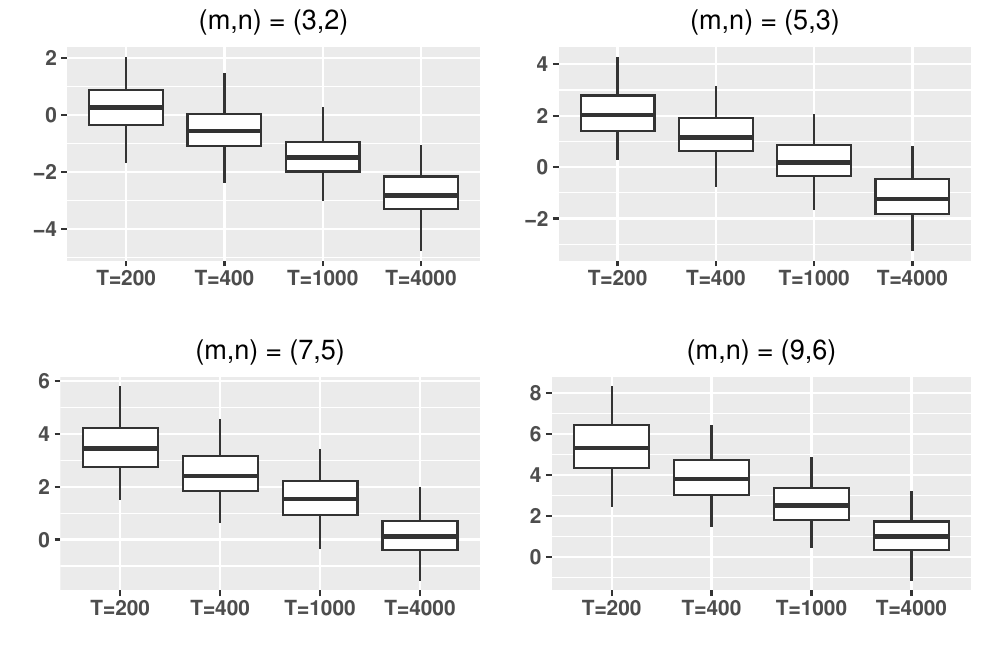}
	\caption{The box plot of the estimation error under Setting II with different  $T$ and $(m,n)$.}
	\label{Plot: LS2}
\end{figure}

For each fixed dimension $(m,n) = (3,2)$, $(5,3)$, $(7,5)$ and $(9,6)$, in Figures \ref{Plot: LS1} and \ref{Plot: LS2}, 
we give the box plot of the estimation  error of autoregressive parameters
\begin{equation*}
    \log \Big( \sum_{i,j=1}^{2}\left\|\widehat{\mathbf{B}}_j\otimes \widehat{\mathbf{A}}_i - \mathbf{B}_j\otimes \mathbf{A}_i \right\|^2\Big)
\end{equation*}
with $T = 200, 400, 1000$, and $4000$. We observe that the estimation error increases as the dimensions $m$ and $n$ grow, due to the inclusion of more items. Besides, compared with Setting I, the more complex structure of $\bSigma$ in Setting II leads to a larger estimation error. Nevertheless, as $T$ increases, the estimation error decreases accordingly across all settings, confirming the consistency of the LSE in Theorem \ref{thm:consistency}.  

\begin{figure}[htbp]
    \centering
    \begin{subfigure}
        \centering
        \includegraphics[width=8.1cm]{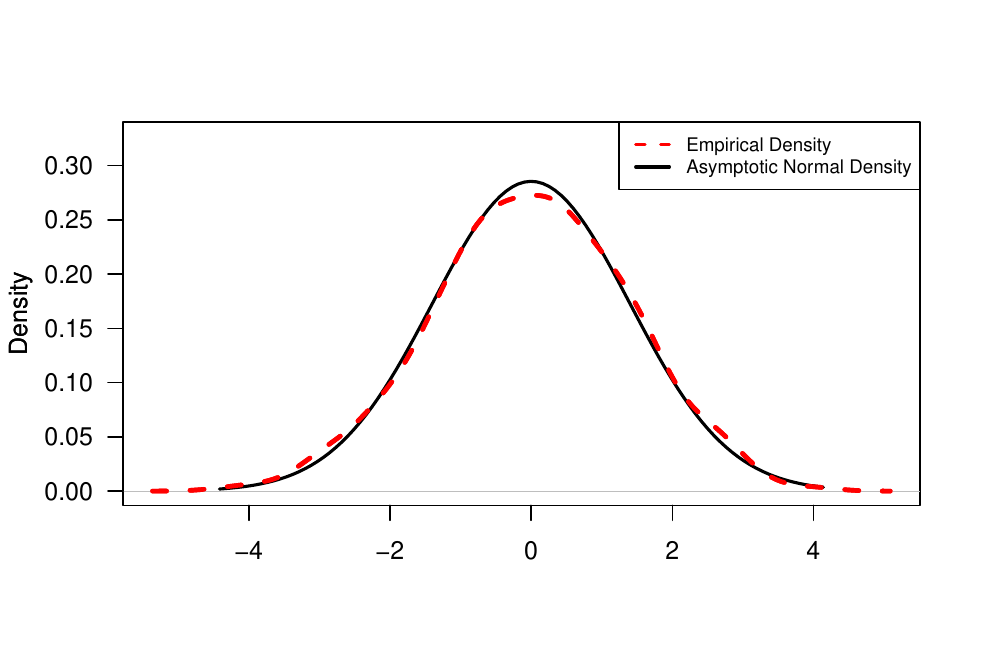}
        %\caption{}
    %\label{fig:Density_A11_Setting_I}
    \end{subfigure}
    %\hfill
    \begin{subfigure}
        \centering
        \includegraphics[width=8.1cm]{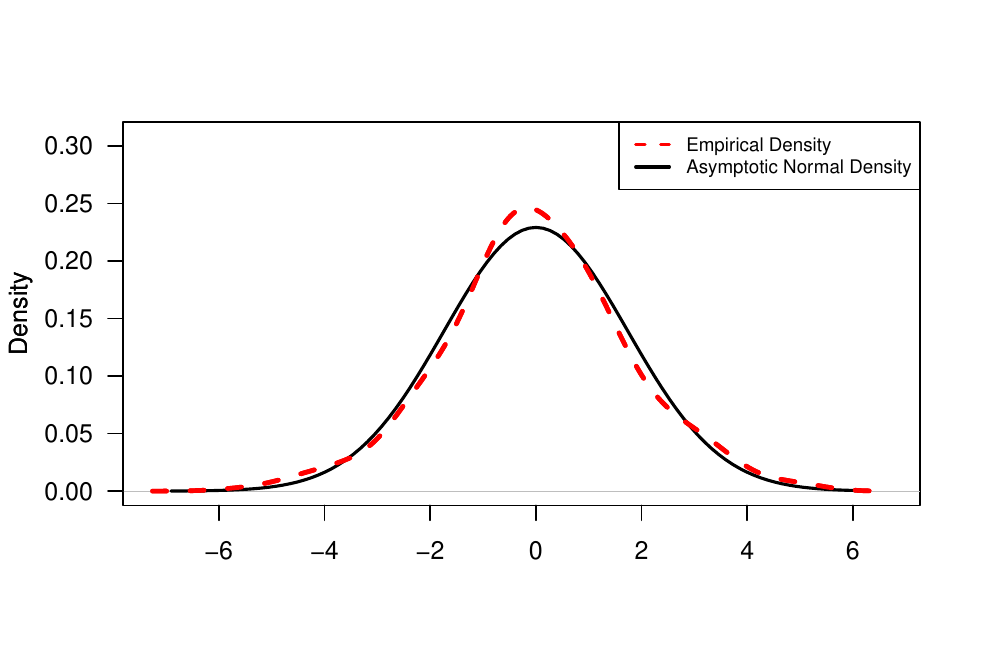}
        %\caption{}
    %\label{fig:Density_A11_Setting_II}
    \end{subfigure}
    \caption{The empirical and theoretical densities of $\sqrt{T}\{(\widehat{\bA}_{1})_{1,1} - (\bA_{1})_{1,1}\}$ under Setting I (left) and Setting II (right) when $T=1000$ and $(m,n) = (3,2)$.}
    \label{Fig:Density_A11}
\end{figure}

To assess the limiting distribution of the autoregressive coefficient matrices, we construct $95\%$ confidence intervals for each element of $\bbeta$, based on the asymptotic normal distribution in Theorem \ref{thm:rate}. 
The empirical coverage probability (ECP) of each true parameter falls within its marginal $95\%$ confidence interval is calculated over 1000 experiments.
Table \ref{Table:CI_AR} summarizes the average ECP across all parameters in $\bbeta$ for different observation lengths $T$ and dimensions $(m, n)$.
\begin{table}[htbp]
  \centering
  \small
  \caption{Average ECP of $95\%$ confidence intervals for autoregressive parameters}
    \begin{tabular}{ccccccccccc}
    \toprule
          &       & \multicolumn{4}{c}{Setting I} &       & \multicolumn{4}{c}{Setting II} \\
\cmidrule{3-11}    $(m,n)$  &       & 200   & 400   & 1000  & 4000  &       & 200   & 400   & 1000  & 4000 \\
    \midrule
    (3,2) &       & 0.936  & 0.941  & 0.944  & 0.948  &       & 0.875  & 0.912  & 0.935  & 0.943  \\
    (5,3) &       & 0.918  & 0.929  & 0.939  & 0.946  &       & 0.848  & 0.895  & 0.922  & 0.937  \\
    (7,5) &       & 0.905  & 0.926  & 0.937  & 0.942  &       & 0.827  & 0.878  & 0.913  & 0.933  \\
    (9,6) &       & 0.893  & 0.922  & 0.934  & 0.941  &       & 0.813  & 0.854  & 0.905  & 0.928  \\
    \bottomrule
    \end{tabular}%
  \label{Table:CI_AR}%
\end{table}%
From this table, we observe that the  ECP increases with the observation length $T$ and approaches the nominal level when $T$ is sufficiently large. Compared with Setting I, the more complex structure of $\bSigma$ in Setting II makes estimation more challenging, thus requiring a larger $T$ to achieve the desired coverage probability.
Besides the confidence interval, we also assess the asymptotic normality by the density function. As an example, consider $(\bA_{1})_{1,1}$, the first element in $\bA_{1}$ when $(m,n) = (3,2)$, and its LSE when $T=1000$. Figure \ref{Fig:Density_A11} shows the empirical  density of $\sqrt{T}\{(\widehat{\bA}_{1})_{1,1} - (\bA_{1})_{1,1}\}$, and the theoretical one from Theorem \ref{thm:rate}. This figure shows that the empirical and theoretical densities are closely aligned in both settings.

Next, we assess the limiting distribution for the threshold parameters $r_0$ and $s_0$.
We construct confidence intervals with different significance levels based on Theorem \ref{thm:asydist}, and summarize the  ECP within $1000$ experiments in Table \ref{Table:Thres_CI_SettingI} for Setting I and Table \ref{Table:Thres_CI_SettingII} for Setting II.
\begin{table}[htbp]
  \centering
  \small
  \caption{ECP of threshold parameters in Setting I}
    \begin{tabular}{ccccccccccccccccc}
    \toprule
          &       &       & \multicolumn{4}{c}{$(m,n) = (3,2)$} &       & \multicolumn{4}{c}{$(m,n) = (5,3)$} &       & \multicolumn{4}{c}{$(m,n) = (9,6)$} \\
\cmidrule{4-7}\cmidrule{9-12}\cmidrule{14-17}          & $1-\alpha$ &       & 200   & 400   & 1000  & 4000  &       & 200   & 400   & 1000  & 4000  &       & 200   & 400   & 1000  & 4000 \\
    \midrule
          & 0.90  &       & 0.81  & 0.86  & 0.89  & 0.90  &       & 0.83  & 0.85  & 0.89  & 0.90  &       & 0.82  & 0.86  & 0.89  & 0.90  \\
    $r_0$    & 0.95  &       & 0.89  & 0.92  & 0.94  & 0.94  &       & 0.90  & 0.93  & 0.94  & 0.94  &       & 0.89  & 0.92  & 0.94  & 0.94  \\
          & 0.99  &       & 0.92  & 0.96  & 0.99  & 0.99  &       & 0.95  & 0.97  & 0.99  & 0.99  &       & 0.93  & 0.97  & 0.98  & 0.99  \\
    \midrule
          & 0.90  &       & 0.82  & 0.86  & 0.90  & 0.90  &       & 0.83  & 0.86  & 0.89  & 0.90  &       & 0.82  & 0.87  & 0.89  & 0.90  \\
    $s_0$    & 0.95  &       & 0.89  & 0.92  & 0.95  & 0.95  &       & 0.90  & 0.92  & 0.94  & 0.95  &       & 0.89  & 0.92  & 0.94  & 0.95  \\
          & 0.99  &       & 0.92  & 0.95  & 0.99  & 0.99  &       & 0.92  & 0.96  & 0.98  & 0.99  &       & 0.93  & 0.97  & 0.98  & 0.99  \\
    \bottomrule
    \end{tabular}%
  \label{Table:Thres_CI_SettingI}%
\end{table}%
\begin{table}[htbp]
  \centering
  \small
  \caption{ECP of threshold parameters in Setting II}
    \begin{tabular}{ccccccccccccccccc}
    \toprule
          &       &       & \multicolumn{4}{c}{$(m,n) = (3,2)$} &       & \multicolumn{4}{c}{$(m,n) = (5,3)$} &       & \multicolumn{4}{c}{$(m,n) = (9,6)$} \\
\cmidrule{4-7}\cmidrule{9-12}\cmidrule{14-17}          & $1-\alpha$ &       & 200   & 400   & 1000  & 4000  &       & 200   & 400   & 1000  & 4000  &       & 200   & 400   & 1000  & 4000 \\
    \midrule
          & 0.90  &       & 0.77  & 0.83  & 0.86  & 0.90  &       & 0.78  & 0.83  & 0.87  & 0.90  &       & 0.78  & 0.84  & 0.87  & 0.91  \\
    $r_0$    & 0.95  &       & 0.85  & 0.89  & 0.92  & 0.94  &       & 0.84  & 0.90  & 0.92  & 0.95  &       & 0.85  & 0.90  & 0.93  & 0.94  \\
          & 0.99  &       & 0.89  & 0.93  & 0.96  & 0.98  &       & 0.89  & 0.93  & 0.96  & 0.99  &       & 0.90  & 0.93  & 0.96  & 0.99  \\
    \midrule
          & 0.90  &       & 0.78  & 0.83  & 0.87  & 0.90  &       & 0.78  & 0.82  & 0.86  & 0.89  &       & 0.79  & 0.83  & 0.87  & 0.91  \\
    $s_0$    & 0.95  &       & 0.84  & 0.88  & 0.91  & 0.95  &       & 0.85  & 0.89  & 0.91  & 0.94  &       & 0.84  & 0.89  & 0.92  & 0.96  \\
          & 0.99  &       & 0.88  & 0.92  & 0.96  & 0.98  &       & 0.89  & 0.92  & 0.96  & 0.99  &       & 0.90  & 0.93  & 0.96  & 0.99  \\
    \bottomrule
    \end{tabular}%
  \label{Table:Thres_CI_SettingII}%
\end{table}%
It is evident that the ECP are quite close to the nominal level, especially when $T$ is large. Furthermore, we verify the asymptotic independence between the two estimated thresholds by utilizing the multivariate independence test based on the empirical copula process proposed by \cite{genest2004test}. By implementing the ``\texttt{indepTest}" function from the ``\texttt{copula}" package in R, we obtain $p$-values of $0.493, 0.387, 0.549$, and $0.523$ for dimensions $(m,n) =(3,2)$, (5,3), (7,5), and $(9,6)$, respectively, with $T=1000$. These results are consistent with the asymptotic independence in Theorem \ref{thm:asydist}.

\section{Real Data Analysis}\label{sec:eg}
\label{sec:Real Data}

In this section, we present two examples to model the nonlinear dynamics of matrix time series. Section \ref{sec:Application1-portfolio} analyzes the dynamics of portfolio returns with threshold effects of size and value, while Section \ref{sec:Application2-Pollution} examines the dynamics of air pollutants with a nonlinear effect of temperature.
To assess the empirical applicability of our model, we include other models for comparison in terms of model fitting and forecasting.
See the details of these comparison models in Table \ref{Table:Model_Comparison}, and their specific forms and implement details in Section S.7 of the Supplementary Material.
\begin{table}[htbp]
  \centering
  \small
  \caption{Model comparison}
    \begin{tabular}{cccc}
    \toprule
          & matrix structure & number of threshold variables   &  number of parameters \\
    \midrule
    2-MART & \ding{51}   & 2        & $2(m^2+n^2) + 2$ \\
    KTMAR & \ding{51}   & 2        & $4(m^2+n^2) + 2$ \\
    SMART & \ding{51}   & 1       & $2(m^2+n^2) + 2$ \\    
    TMAR & \ding{51}   & 1        & $2(m^2+n^2) + 1$ \\
    TMAR(3)  & \ding{51}   & 1        & $3(m^2+n^2) + 2$ \\
    MAR   & \ding{51}   & 0        & $m^2+n^2$ \\
    VAR   & \ding{55}    & 0         & $m^2n^2$ \\
    RRVAR & \ding{55}    & 0         & $2mnk$ \\
    \bottomrule
    \end{tabular}%
   \label{Table:Model_Comparison}%
  \begin{tablenotes}
  \footnotesize
  \item Note: $(m,n)$ is the dimension of  matrix-valued time series $\bX_t$ and $k$ is the rank in (S.7.4).
\end{tablenotes}
\end{table}%

Specifically, in addition to our 2-MART model, we also consider a multiple-threshold-variable matrix autoregressive model (KTMAR) in (S.7.1), which generalizes the KTAR model \citep{zhang2024least} to the matrix setting.
To compare, the KTMAR model also involves two threshold variables and four regimes, but assuming that both row and column coefficients vary across four regimes, introducing a large number of parameters in the matrix setting.

We also include models with a single threshold variable. 
The Softened Matrix AutoRegressive with Threshold model (SMART) in \eqref{Multiple-TMAR}, a special case of our 2-MART model, is advantageous when the row and column share the same threshold variable, as in the pollutant analysis. However, it may be less effective when they do not, as in the portfolio analysis. 
The threshold matrix autoregressive model (TMAR) in \eqref{TMAR} with two regimes by default, and its generalization with three regimes (TMAR(3)) in (S.7.2), are also considered.
These models assume the same nonlinear effects for rows and columns, while the SMART model allows for heterogeneity.
Additionally, the SMART model considers three regimes, including an intermediate regime, offering more flexibility than the TMAR model and requiring fewer parameters than the TMAR(3) model. 
Due to this advantage, the SMART model is specifically fitted in the pollutant analysis. 
For these threshold models with a single threshold variable, the model is fitted with either $z_{t-1}$ or $w_{t-1}$ as the threshold variable, and the one with better performance in model fitting is reported.

Linear models are also  considered, including the matrix autoregressive model (MAR) in \eqref{MAR},  vector autoregressive model (VAR) in (S.7.3) and reduced rank vector autoregressive model (RRVAR) in (S.7.4).
The VAR model studies the vectorized matrix, hence introducing the most parameters and potentially leading to model identification and fitting problems.
The RRVAR model mitigates this issue to some extent with a low rank structure.
Specifically, the MAR model is estimated by the function ``\texttt{tenAR.est}" in the R package ``\texttt{tensorTS}", and the RRVAR model by the function ``\texttt{rrr.fit}" in the R package ``\texttt{rrpack}".

\subsection{Analysis of Portfolio Returns}
\label{sec:Application1-portfolio}
\begin{figure}[!htbp]
	\centering	\includegraphics[height=2.1cm, width=14cm]{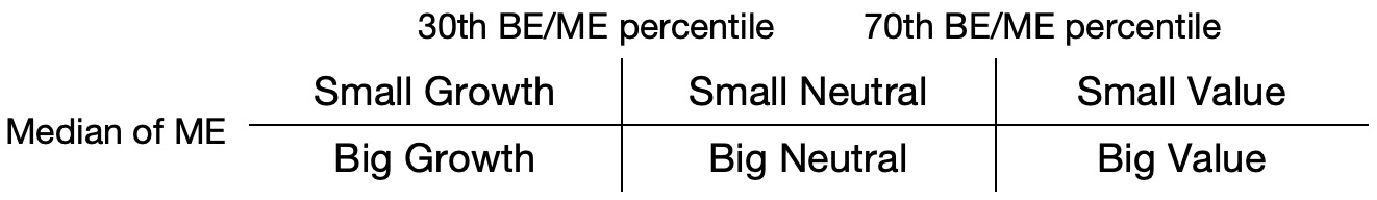}
	\caption{Portfolio construction}
	\label{Plot: Factor}
\end{figure}

We now revisit the portfolio returns example mentioned in Section \ref{sec:intro}. Specifically, we group a universe of stocks into $6$ portfolios based on $2$ levels of size (market equity, ME) and $3$ levels of value (ratio of book equity to market equity, BE/ME).
The average value weighted returns within each group are calculated, forming a $2 \times 3$ matrix-valued time series $\X_t$.
See Figure \ref{Plot: Factor} for an illustration. We analyze their standardized weekly returns from May 24, 2002, to January 26, 2024, with $T=1130$ and $m=2$, $n=3$. The data source and further details can be found on  Kenneth R. French website.\footnote{ http://mba.tuck.dartmouth.edu/pages/faculty/ken.french/data\_library.html}

To account for the nonlinear characteristics in the financial market, a threshold model is constructed to examine the interaction of size and value factors, providing insights into stock returns and improving investment strategies.
The size effect and value effect are considered in the famous Fama--French three-factor model. Specifically, the small minus big (SMB) factor represents the size effect, referring to the tendency for smaller companies (by market capitalization) to outperform larger ones in terms of returns. The high minus low (HML) factor captures the value effect, where value stocks with high BE/ME ratios outperform growth stocks with low BE/ME ratios.
Different SMB and HML factors indicate varying preferences for stocks in the market, potentially resulting in different dynamics of stock returns.
Hence,  they are chosen as the row and column threshold variables in modeling $\X_t$, respectively. 
Specifically, they are calculated as
\begin{align*}
    \text{SMB: }z_{t} = \frac{1}{3}\sum_{j=1}^3\{ (\mathbf{X}_t)_{1, j} - (\mathbf{X}_t)_{2, j}\}, \quad
    \text{HML: }w_{t} = \frac{1}{2}\sum_{i=1}^2\{(\mathbf{X}_t)_{i, 3} - (\mathbf{X}_t)_{i, 1}\}.
\end{align*}

We estimate the 2-MART model (\ref{Two-TMAR}) for $\X_t$.
The estimated threshold parameters for row and column are $\widehat{r} = -0.278$ and $\widehat{s} = -0.029$.
The estimated coefficient matrices are shown below:
\begin{align}\label{eq:coef}
\begin{split}
    \A_1 &=
    \begin{pmatrix}
        0.394 & -0.328\\
        0.648 & -0.563
    \end{pmatrix} ,\quad
    \A_2 =
    \begin{pmatrix}
         -0.199 & -0.452 \\
         0.091 & -0.865 \\
    \end{pmatrix},\\
    \B_1 &=
    \begin{pmatrix}
     0.104 & 0.438 & 0.159  \\
    -0.035 & 0.091 & 0.468  \\
    0.008 & -0.428 & 0.947 \\
    \end{pmatrix} ,\quad
    \B_2 =
    \begin{pmatrix}
    0.175 & 0.102 & 0.005 \\
     0.106 & 0.091 & 0.166 \\
     0.089 & 0.165 & 0.196 \\
    \end{pmatrix}.
    \end{split}
\end{align}

The model provided insights into stock returns dynamics, revealing that each regime indicates a dominance or preference for certain types of stocks in the market. By understanding which regime is currently predominant, investors can better align their investment strategies with the prevailing market conditions, potentially enhancing returns and managing risks more effectively. Specifically, the four regimes identified are:
\begin{itemize}
\item  Regime 1 (Preference for Small and Growth stocks with $\A_1$, $\B_1$ and $T_{11} = 479$):  The coefficient matrices $\mathbf{A}_1$ and $\mathbf{B}_1$ reflect strong autoregressive effects and interactions within this group. For instance, the positive elements (or row sums) in $\mathbf{A}_1$ suggest that returns within this regime are more self-reinforcing, which is typical of high-growth stocks reacting strongly to their own past performance. The predominance of this regime suggests potential investor confidence in economic growth.
\item Regime 2 (Preference for Small and Value stocks with $\A_1$, $\B_2$ and $T_{12} = 415$):  
The preferred firms are often overlooked by the market but may possess strong fundamentals not fully reflected in their stock prices. The matrix $\mathbf{B}_2$ indicates positive coefficients, suggesting strong future performance for these undervalued stocks as their true value is recognized by the market. The prevalence of this regime could indicate a market correction where investors are looking for bargains among small caps, possibly during uncertain economic times.
\item Regime 3 (Preference for Big and Growth stocks with $\A_2$, $\B_1$ and $T_{21} = 131$): The preferred firms consist of larger and well-established companies that are often perceived as safer investments. These companies tend to have stable earnings, and their low book-to-market ratios suggest a growth stock profile. The matrix $\mathbf{B}_1$ for this regime indicates moderate interactions among stocks, reflecting the stability and lower volatility characteristic of large growth stocks. The negative entries in $\A_2$ suggest some corrective mechanisms in place, which is typical for large companies adjusting their values based on market conditions. The dominance of this regime suggests a preference for safety and quality, with investors favoring large, stable companies.
\item Regime 4 (Preference for Big and Value stocks with $\A_2$, $\B_2$ and $T_{22} = 105$): 
The preferred stocks consist of larger companies that are potentially undervalued. These might be mature, stable companies with solid fundamentals that are not fully priced into their market valuations.  The positive entries in matrix $\B_2$ indicates significant gains as the true value of these undervalued large companies is recognized by the market. The negative entries in $\A_2$ suggest a corrective mechanism where past performance may negatively impact future performance, highlighting the undervaluation adjustments. This regime indicates a market environment where value investing is favored, reflecting investors' caution and search for stable, reliable investments at reasonable prices.
\end{itemize}

Table~\ref{Table_MSPE} presents the number of parameters $p$ (including autoregressive coefficients and thresholds)
 and the mean squared prediction error (MSPE) values for each model.
Specifically, let $t \in \{1051,1052,\ldots,1130\}$ represent the rolling forecasting set, with a training sample size of $1050$.
For each $t$ in the set, 
data from $t-1050$ to $t-1$ is used to fit the model, which then generates a one-step forecast, denoted as $\widehat{\bX}_{t}$. 
The MSPE is computed as $\sum_{t=1051}^{1130}\|\widehat{\bX}_{t} - \bX_{t} \|^2/80$, 
and is reported for our 2-MART model along with other models in Table~\ref{Table_MSPE}.
From Table~\ref{Table_MSPE}, we observe that all the matrix models outperform the vector models by a substantial margin, with our 2-MART model achieving the lowest MSPE value while maintaining a reasonable number of parameters.  
This result demonstrates the applicability and advantage of our 2-MART model.
\begin{table}[!htbp]
  \centering
  \small
  \caption{Mean squared prediction error of different models for the portfolio returns data}
    \begin{tabular}{ccccccccc}
    \toprule
          & 2-MART &  KTMAR & SMART  & TMAR & TMAR(3)  & MAR   & VAR   & RRVAR \\
    \midrule
    $p$     & 28    & 54    & 28    & 27    & 40    & 13    & 36    & 24 \\
    MSPE   & 1.71 &  1.89 &  1.75   & 1.78 & 1.83 & 1.81 & 2.61 & 2.38 \\
    \bottomrule
    \end{tabular}%
  \label{Table_MSPE}%
\end{table}%

\subsection{Analysis of Air Pollution Data}
\label{sec:Application2-Pollution}
In the second example, we examine the impact of meteorology on the dispersion and deposition of atmospheric pollutants.
Temperature and related quantities, such as temperature perturbation, in particular, exert significant nonlinear effects on multiple pollutants, influencing their concentration and distribution in complex ways \citep{aw2003evaluating,yang2021nonlinear}. 
Understanding these interactions is essential for developing effective pollution control strategies and mitigating the severe health risks associated with poor air quality.
To address this, we build a SMART model \eqref{Multiple-TMAR} to depict the nonlinear dynamics of multiple pollutants across various sites.

The data consists of
daily measurements of three pollutants: the suspended particulate matter with a diameter of $2.5$ micrometers or less ($\text{PM}_{2.5}$), the suspended particulate matter with a diameter of $10$ micrometers or less ($\text{PM}_{10}$) and the nitrogen dioxide ($\text{NO}_2$). The data spans from January 1, 2017, to December 31, 2019, and was collected from four air quality monitoring stations in Seoul. This results in a $4 \times 3$ matrix-valued time series $\bX_t$, where rows correspond to different monitoring stations and columns represent different pollutants, with the observation length of $T = 1095$. The series has been pre-standardized. 
For the threshold variable $z_t$, we consider the temperature perturbation from the previous day, i.e., $z_t = u_{t} - u_{t-1}$, where $u_t$ is the temperature in Seoul on date $t$ (averaged across regions within the city). The data is collected from the website of the National Climatic Data Center (NCDC) \footnote{https://www.ncei.noaa.gov/cdo-web/datasets} and Kaggle \footnote{https://www.kaggle.com/datasets/bappekim/air-pollution-in-seoul/data}.

The estimated threshold parameters for row and column are $\widehat{r} = -4.05$ and $\widehat{s} = 0.95$, and the length of observations in $3$ regimes are $\{T_{11} = 148, T_{21} = 427, T_{22} = 519\}$. The estimated autoregressive coefficient matrices are shown in (\ref{eq:coef_Pollution}).

\begin{align}\label{eq:coef_Pollution}
\begin{split}
   \A_1 &=
    \begin{pmatrix}
     0.165 &	0.182 &	-0.034 & 0.253   \\
    -0.357 &	0.319 &	-0.007 & 0.257   \\
    -0.167 &	0.117 &	0.014 &	-0.146  \\
    -0.271 &	-0.167 &	-0.013 &	0.441 
    \end{pmatrix} ,\quad
    \A_2 =
    \begin{pmatrix}
    0.320 &	0.284 &	0.169 &	0.127  \\
    0.042 &	0.658 &	0.147 &	0.049  \\
    0.068 &	0.307 &	0.384 &	0.129 \\
    0.165 &	0.080 &	0.126 &	0.404 
    \end{pmatrix},\\
    \B_1 &=
    \begin{pmatrix}
    0.352 &	0.158 &	0.291    \\
    0.013 &	0.459 &	0.282   \\
   -0.002 & 0.104 &	0.216  \\
    \end{pmatrix} ,\quad
    \B_2 =
    \begin{pmatrix}
     0.592 &	0.186 &	0.314   \\
     0.054 &	0.630 &	0.369  \\
     0.085 &	0.265 &	0.511  \\
    \end{pmatrix}.
    \end{split}
\end{align}

Furthermore, we evaluate the out-of-sample forecasting performance of these models, which is crucial for effective pollutant prevention and control.
Let $t \in \{1001,1002,\ldots,1095\}$ be a rolling forecasting set with a  training sample size of 1000.
For each $t$ in the set, 
data from $t-1000$ to $t-1$ is used to fit the model, based on which a one step forecast $\widehat{\bX}_{t}$ is generated.  
The mean squared prediction error (MSPE), calculated as $\sum_{t=1001}^{1095}\|\widehat{\bX}_{t} - \bX_{t} \|^2/95$, 
for our SMART model alongside other models in Table \ref{Table_Pollution_AIC}; 
\begin{table}[htbp]
  \centering
  \small
  \caption{Mean squared prediction error for different models}
    \begin{tabular}{ccccccc}
    \toprule
    & SMART & TMAR & TMAR(3)  & MAR   & VAR   & RRVAR \\
    \midrule
    $p$     & 52    & 51    & 77    & 25    & 144   & 126 \\
    MSPE  & 9.84  &  10.25 & 11.17 & 10.89 & 16.86 & 15.28 \\
    \bottomrule
    \end{tabular}%
  \label{Table_Pollution_AIC}%
\end{table}%
see the details of these comparison models in Table \ref{Table:Model_Comparison}. Again, we note that all the matrix models outperform the vector models by a substantial margin.
Notably, our model demonstrates superior forecasting accuracy, further validating its effectiveness and robustness.
Compared with the TMAR model, the SMART model introduces only one additional threshold parameter, allowing for distinct threshold effects for rows and columns, which significantly improves the performance of model fitting and prediction. 
This result highlights the importance of incorporating a two-way threshold structure.
It is noteworthy that it has  improved the forecast accuracy over the linear MAR model by about 10\%. 
It shows the nonlinear effect of temperature perturbation on pollutant dynamics, effectively delineating three distinct regimes with a softened threshold transition mechanism.

\section{Conclusion and Discussion}
\label{sec:Conclusion}
In this article, we conduct an in-depth study of a nonlinear matrix process and propose an innovative two-way matrix autoregressive model with thresholds. This model captures regime-shifting behaviors by examining 
row-wise and column-wise threshold effects separately, enabling the study of the nonlinear dynamics within the rows and the columns, and hence providing a clear and comprehensive understanding of the underlying matrix processes.
We develop methods to estimate the threshold and matrix autoregressive parameters, and we establish the asymptotic properties of the estimators under certain conditions. 
The two-way nature of the model significantly reduces the number of model parameters and computational costs, enhancing both model identification and fitting.
Consequently, our model strikes a balance between model simplicity and the ability to depict nonlinear dynamics.

Our model presents a novel approach for analyzing the nonlinear characteristics of complex structured time series. 
Potential directions for future research include multiple regimes generalization for both rows and columns, threshold matrix model with heavy-tailed data, threshold effect (numbers of threshold variables and parameters) testing in the matrix setting, among others.
In a broader context, exploring nonlinearity could be combined with matrix factor models and matrix change-point detection.
We recognize the potential to study the nonlinear matrix factor model, which could significantly enhance the applicability of this model.
Investigating scenarios where both change points and nonlinear effects are present is also of great interest.
We leave these compelling topics as future directions.

\section*{Supplementary Material}
The \emph{Supplementary Material} includes all proofs, model generalizations, and more empirical implementation details.

\linespread{1.3}\selectfont
\bibliographystyle{dcu}
\bibliography{ref}

\end{document}